\documentclass[aps,graphicx,twocolumn]{revtex4}
\usepackage{amsmath}
\usepackage{amscd}
\usepackage{graphicx}
\usepackage{multirow}
\begin{document}

\title{Analysis of $N$-qubit perfect controlled teleportation schemes from the controller's point of view}
\author{ Xi-Han Li$^{1,2}$\footnote{
Email address: xihanlicqu@gmail.com}, Shohini Ghose$^{1,3}$\footnote{
Email address: sghose@wlu.ca}}
\address{$^1$ Department of Physics and Computer Science, Wilfrid Laurier University, Waterloo, Canada \\$^2$Department of Physics, Chongqing University,
Chongqing, China\\
$^3$ Institute for Quantum Computing, University of Waterloo, Canada}

\date{\today }
\begin{abstract}
We quantitatively analyze and evaluate the controller's power in $N$-qubit controlled teleportation schemes. We calculate the minimum control power required to ensure the controller's authority such that the teleportation fidelity without the controller's permission is no more than the classical bound. We revisit several typical controlled teleportation schemes from the controller's point of view and evaluate the control power in these schemes. We find that for teleporting arbitrary $N$-qubit states, each controller should control at least $N$ bits of useful information to ensure his/her authority over the protocol. We also discuss the general rules that must be satisfied by controlled teleportation schemes to ensure both teleportation fidelity and control power.
\end{abstract}
\maketitle

\section{Introduction}
Entanglement is a significant resource in many quantum information processing protocols including quantum dense coding \cite{dense1,dense2}, quantum key distribution \cite{qkd1,qkd2,qkd3},
quantum secret sharing \cite{qss1,qss2,qss3},
quantum secure direct communication \cite{qsdc1,qsdc2} and quantum computation\cite{qc1,qc2}.
An important application of entanglement that has been widely studied is quantum teleportation \cite{tele}.
An unknown quantum state can be teleported from one site to another via previously shared entanglement assisted by classical communications and local operations. The first quantum teleportation scheme proposed in 1993 showed how a two-qubit maximally entangled Bell state can be employed as the quantum channel to transfer an arbitrary single-qubit state \cite{tele}. Since then, quantum teleportation has attracted much attention. Teleportation of multi-qubit states or $d$-dimension states were proposed and teleportation experiments were demonstrated in the lab \cite{t1,t2,t3,t4,t5,t6,t7}.

In this paper, we focus on a variation of quantum teleportation, called controlled teleportation (CT) \cite{ct}. In CT schemes, the teleportation procedure between Alice and Bob is controlled by a third party, Charlie.  Controlled teleportation has interesting applications in the context of networked quantum information processing and cryptographic conferencing \cite{use1,use2,use3,use4}. Compared with the standard teleportation procedure, CT schemes should consider the controller's role in addition to the teleportation itself.  In other words, the controller's authority should be ensured as well as the fidelity of teleportation of the final state. However, existing CT schemes have not analysed the controller's authority in a quantitative way. Here, we present a quantifiable measure of the controller's power in $N$-qubit teleportation. We identify a lower bound on the control power that CT schemes must meet in order to ensure the controller's authority by requiring that the teleportation fidelity without the permission of the controller should be minimized in order to maximize the controller's power.  We apply our measure to evaluate the performance of several existing CT protocols.

The first CT scheme utilized the three-qubit Greenberger-Horne-Zeilinger (GHZ) state as a quantum channel to teleport a single-qubit state under the control of one agent. We henceforth call this the GHZ scheme for simplicity. Later, Deng \emph{et al}. presented a method for symmetric multiparty-controlled teleportation of an arbitrary two-particle entangled state by using two GHZ states (2-GHZ scheme) \cite{deng}. In 2006, Li \emph{et al}. proposed an efficient symmetric multiparty quantum state sharing
scheme, which can also be used for controlled teleportation  of an arbitrary $N$-qubit state via $N$ GHZ states ($N$-GHZ scheme) \cite{li}. However, this scheme requires considerable auxiliary qubit resources when $N$ is significantly large.  In 2004, Yang \emph{et al}. proposed efficient many-party CT protocols to teleport $N$-qubit product states \cite{yang1, yang2}. In 2007, Man \emph{et al}. constructed a genuine $(2N+1)$-qubit entangled channel to perform controlled teleportation of an arbitrary $N$-qubit state controlled by one agent \cite{man1} and then generalized it to $M$ controllers via $M$ GHZ states and $(N-M)$ Einstein-Podolsky-Rosen (EPR) pairs \cite{man2}.
Several other CT schemes have been proposed, which differ in the quantum channel or the states be teleported \cite{gao,ct2,ct5,ct6}. In particular, Gao \emph{et al.} pointed out that there exist partially entangled three-qubit states which can be used for perfect controlled teleportation (PCT) of a single qubit - i.e. - with unit success probability and fidelity~\cite{gao} .
Since it not easy to prepare and maintain maximal entanglement in practice, their scheme has both theoretical and practical significance.


Recently, we presented a quantity for measuring the controller's power in single-qubit controlled teleportation and identified a reasonable criterion for evaluating whether a particular quantum channel is suitable for controlled teleportation of  a single-qubit state \cite{we}. We analyzed the controller's power in CT via three-qubit entangled channels and showed that three-qubit partially entangled channels are unsuitable for perfect controlled teleportation of arbitrary single-qubit quantum states because they do not ensure the controller's power. In this paper, we generalize the case to discuss the controller's power in the controlled teleportation of $N$-qubit states. We establish a lower bound on the controller's power that will ensure the controller's authority in a perfect controlled teleportation scheme. Then we revisit several typical CT schemes with our criterion. We compare the control power in two classes of CT schemes for teleporting $N$-qubit states. In one class, each controller is assigned one qubit and only single-qubit measurements are required, while in the other class each controller owns $N$ qubits. We find that there is a trade-off between the controller's power and the quantum resources consumed. Based on our criterion, each controller should control at least $N$ bits of useful information to ensure his/her power. Finally, we give a summary of the general rules that a controlled teleportation scheme must satisfy in order to ensure both perfect teleportation as well as the controller's authority.

\section{controller's power in controlled teleportation schemes}
We first review our definition of controller's power in controlled teleportation \cite{we}. Suppose in a controlled teleportation protocol, the sender, receiver and the controller are Alice, Bob and Charlie, respectively. The state to be teleported from Alice and Bob is unknown to all of them, denoted by $\vert \varphi\rangle_X$. To teleport this state under the control of Charlie, the three parties share an entangled channel in advance. First, the sender performs  a joint measurement on her entangled particle and the unknown state and sends Bob the measurement outcome via a classical channel. Then, if the controller Charlie, wishes the teleportation to be executed, he measures his own particle and sends the result to Bob. With these measurement results, Bob can rotate his state back to the input state to be teleported via appropriate unitary operations. The controller's power is determined by how much information Bob can obtain without the controller's help. If Charlie does not disclose his measurement results, Bob's state is a mixed state $\rho_B$ even with Alice's results. The density matrix can be computed by $\rho_B=tr_C(\vert \psi\rangle_{BC}\langle \psi \vert)$, where $\vert \psi\rangle_{BC}$ is the state of Charlie and Bob's qubits after Alice's measurement.
Then we can calculate the non-conditioned fidelity (NCF) of Bob's state, the fidelity without Charlie's help as \cite{ct}
\begin{eqnarray}
f=\langle \varphi \vert \rho_B \vert \varphi\rangle.
\end{eqnarray}
Usually, the fidelity depends on the target state and the average fidelity $\bar{f}$ can be obtained by averaging over all input states. If the controller cooperates, the conditioned fidelity (CF) of Bob's state is unity since we focus on perfect controlled teleportation, i.e., Bob can recover the original state deterministically with Charlie's help. Therefore, the controller's power $CP$ can be defined as the difference between the CF and the NCF, the more the better.
\begin{eqnarray}
CP=1-\bar{f}.
\end{eqnarray}

In above description, the state to be teleported can be either a single qubit or an $N$-qubit state. To ensure the controller's authority, Bob's NCF should be as small as possible - i.e, the teleportation fidelity without the permission and participation of the controller should be minimized. For teleporting a single-qubit state, the minimum fidelity is $1/2$, which corresponds to a random guess. Therefore, the maximal control power for single-qubit CT is $CP^{(1)}_{max}=1/2$. The classical limit of fidelity for a  single-qubit state is $2/3$, which is the best fidelity via classical teleportation \cite{2/3,2/3+}. Then the NCF should be no more than the classical limit $2/3$, and hence the lower bound on the control power is $CP^{(1)} \geq 1/3$.

Now we look at CT schemes for teleporting an $N$-qubit state with $M$ controllers. The $N$-qubit arbitrary state can be written as
\begin{eqnarray}
\vert \varphi\rangle_{X}=\sum^1_{i_n=0} \alpha_{i_1,i_2,...i_N} \vert i_1\rangle_{X_1} \vert i_2\rangle_{X_2}...\vert i_N\rangle_{X_N}
\end{eqnarray}
Here $\sum^1_{i_n=0} \vert \alpha_{i_1,i_2,...i_N}\vert^2=1$ and $X$ denotes the $N$ qubits.
We can generalize the method to calculate the NCF of arbitrary $N$-qubit states in different CT schemes. In order to compute the $m$th controller's control power, we let Alice and the other controllers perform their measurements and obtain the collapsed state $\vert \psi\rangle_{C_mB}$. Here $B$ denotes the $N$ qubits with Bob and $C_m$ denotes the particles (can also be one particle) held by the $m$th controller who is not participating. Then, tracing over the  $C_m$ we get the density matrix $\rho_B$ and the NCF without the permission of the $m$th controller,  $f=\langle \varphi \vert \rho_B \vert \varphi\rangle$.

For an $N$-qubit input state, the minimal fidelity by guessing is $1/2^N$, and the classical limit is \cite{cl1,cl2,cl3}
\begin{eqnarray}
F_{cl}=\frac{2}{1+2^N}.
\end{eqnarray}
This means that for perfect controlled teleportation of an $N$-qubit state, the controller's power should be $CP^{(N)}\geq \frac{2^N-1}{2^N+1}$.

\section{Revisiting controlled teleportation schemes from the controller's view point}
We now analyze existing CT schemes to test if the control power in these schemes meets our lower bound.
Let's test the 2-GHZ scheme first\cite{deng}. In this scheme for teleporting two-qubit states, two GHZ states are employed as the quantum channel. The sender Alice performs two Bell state measurements and the controller Charlie makes one Bell state measurement. Without Charlie's measurement results, Bob's qubits are in a mixed state of four possible pure states that can be rotated to
\begin{eqnarray}
\rho_B=\frac{1}{4}\vert \varphi\rangle\langle\varphi\vert +\frac{1}{4}\sum^3_{i=1}\vert \phi_i\rangle\langle\phi_i\vert,
\end{eqnarray}
where $\vert \varphi\rangle$ is the desired state and $\vert \phi_i\rangle$ denotes other possible states.
It is not difficult to calculate the control power by averaging over all input states.
\begin{eqnarray}
CP^{(2)}_{2-GHZ}=1-\bar{f}_{2-GHZ}=3/5,
\end{eqnarray}
This is equal to the lower limit of control power for two-qubit states, $CP^{(2)}=3/5$. This means that Deng's protocol is an acceptable CT protocol from the controller's point of view, since the control power is not less than our lower bound.

Although the 2-GHZ scheme can realize perfect CT while ensuring the controller's power, it requires the controller to perform Bell-state measurements.  In the $N$-GHZ scheme \cite{li}, the controller only needs to make single-qubit product measurements. Each controller owns $N$ qubits in the $N$-GHZ scheme, and the corresponding measurement results have $N$ bits of information. Therefore, without one controller's information, Bob's state is a mixed state of $N$ possible pure states, which can be rotated to the following one with the other agents' information:
\begin{eqnarray}
\rho_B=\frac{1}{2^N}\vert \varphi\rangle\langle\varphi\vert +\frac{1}{2^N}\sum^{2^N-1}_{i=1}\vert \phi_i\rangle\langle\phi_i\vert.
\end{eqnarray}
 It is not difficult to verify that each controller has the same control power $CP^{(N)}_{N-GHZ}=\frac{2^N-1}{2^N+1}$ which confirms the suitability of this  for scheme a CT task. However, this scheme requires a large number of auxiliary qubits and measurements, which increases the cost of the protocol. Therefore, CT schemes in which each controller possesses only one qubit were proposed \cite{yang1,yang2,man1,man2}. We now analyze these proposals from the controller's point of view.

We can discuss Ref.\cite{yang1} and \cite{yang2} together since they involve the same basic principle. The state to be teleported is the multiqubit product state
\begin{eqnarray}
\vert \varphi'\rangle_X=\prod^N_{n=1}(\alpha_n\vert 0\rangle_{X_n}+\beta_n \vert 1 \rangle_{X_n}).
\end{eqnarray}
The quantum channel is
\begin{eqnarray}
\prod^N_{n=1}\vert EPR\rangle_{A_nB_n}\otimes\vert GHZ\rangle_+ +\prod^N_{n=1}\vert \widetilde{EPR}\rangle_{A_nB_n}\otimes\vert GHZ\rangle_-\nonumber\\
\end{eqnarray}
where $\vert EPR\rangle_{A_nB_n}=\frac{1}{\sqrt{2}}(\vert 00\rangle+\vert 11 \rangle)_{A_nB_n}$ and $\vert \widetilde{EPR}\rangle_{A_nB_n}=\frac{1}{\sqrt{2}}(\vert 00\rangle-\vert 11 \rangle)_{A_nB_n}$ in Ref.\cite{yang1} and $\vert \widetilde{EPR}\rangle_{A_nB_n}=\frac{1}{\sqrt{2}}(\vert 01\rangle-\vert 10 \rangle)_{A_nB_n}$ in Ref.\cite{yang2}. $\vert GHZ \rangle_{\pm}=\frac{1}{\sqrt{2}}(\vert 0\rangle_C^{\otimes M}\vert 0\rangle_A\pm\vert 1 \rangle_C^{\otimes M}\vert 1\rangle_A)$ are $(M+1)$-qubit GHZ states. To calculate the $m$th controller's control power, we let the other controllers and Alice perform their measurements so that we are left with a state composed of $C_m$ and $B$. Then we trace over $C_m$ to get the density matrix of $B$ to compute the NCF. It is not difficult to verify that the NCF is always larger then $1/2$, which is definitely larger then the classical limit for teleporting an $N$-qubit state. This implies inadequate control power. However, since the state to be teleported is an $N$-qubit product state, we can compute the controller's control power for each qubit $B_n$ by further tracing over Bob's other qubits. Then the average NCF is $2/3$ which does meet the lower bound for control power.
If we use the quantum channel to teleport arbitrary $N$-qubit states instead of product states,  the average NCF for Bob's $N$-qubit state without the $m$th controller's permission is
\begin{eqnarray}
\bar{f}_{Yang}= \frac{2^{N-1}+1}{2^{N}+1}, \label{yang}
\end{eqnarray}
which is always larger than the classical limit $\frac{2}{1+2^N}$ when $N>2$. This implies the quantum channels used in Yang's schemes are not suitable for teleporting arbitrary $N$-qubit states from the controller's point of view.

Another class of CT protocols in which each controller has one qubit was proposed by Man \emph{et al}.\cite{man1,man2}. The quantum channel is composed of $M$ GHZ states and $(N-M)$ EPR pairs $(M\geq 1)$. These schemes were designed for teleporting arbitrary unknown $N$-qubit entangled states.  If a single controller does not participate in the CT then Bob's state is a mixture of two pure states with equal probability. Therefore, it is similar to Yang's protocols that the average NCF is definitely larger than $1/2$ which is larger than the classical limit when $N>2$. This means a lack of adequate control power. If we use this quantum channel for teleporting arbitrary $N$-qubit states, the average NCF is the same as Eq. (\ref{yang}), which means these schemes are also unsuitable for CT from the controller's point of view. Moreover, Man's schemes can also not be used for $N$-qubit product states. In that case, each controller can only control one qubit. And if $M<N$, there are $(N-M)$ particles uncontrolled.

To sum up, although the schemes in which each controller only performs single-qubit operations are economical from the resource point of view, they cannot meet the minimum requirement for the controller's authority for teleporting arbitrary $N$-qubit states. Compared with the $N$-GHZ scheme \cite{li}, we find there is a trade off between the resources consumed and the control power. To ensure the controller's power, more quantum resources are required.

\section{discussion and summary}

In conclusion, for a general CT protocol that teleports arbitrary multiqubit states via many controlling agents, certain criteria need to be satisfied. Firstly, each qubit should be controlled. In Ref.\cite{man1,man2}, the controllers' measurement results only has impact on $M$ of Bob's $N$ qubits. Therefore, Bob can get $N-M$ qubits of information without the controller's help. Secondly, each controller should have the same power in the $(m,m)$-threshold CT scheme. Thirdly, the controller's power should be restricted to some range in order to ensure his/her authority. In this paper, we use the classical limit to restrict the controller's power, which we think is a reasonable bound to prevent the receiver from obtaining any non-classical fidelity without the controller's permission.

Based on our criteria, we can easily estimate the number of qubits each controller should possess in order to adequately control the teleportation of arbitrary $N$-qubit states. Suppose the controller possesses $W$ qubits. Then he/she has $2^W$ measurement results at most. For a maximally entangled quantum channel in which each result has equal probability $1/2^W$, Bob can get the following density matrix without Charlie's measurement results,
\begin{eqnarray}
\rho_B=\frac{1}{2^W}\vert \varphi \rangle\langle \varphi \vert +\frac{1}{2^W}\sum^{2^W-1}_{i=1}\vert \phi_i \rangle\langle \phi_i \vert.
\end{eqnarray}
Therefore, the NCF is always larger than $\frac{1}{2^W}$. When Charlie has $N-1$ qubits, we have
\begin{eqnarray}
\bar{f}>\frac{1}{2^{N-1}}>\frac{2}{2^N+1},
\end{eqnarray}
which means Bob can achieve a  better-than-classical fidelity if Charlie only has $N-1$ qubits in hand. Therefore, for teleporting $N$-qubit states, each controller should have at least $N$ qubits to ensure the minimum control power. It should be emphasized that the $N$ qubits is only a necessary condition but not a sufficient one. The control power depends on the quantum channel and strategy. If we do not limit to a two-level system, the controller can use a qudit and each one could control $N$ bits information for teleporting $N$-qubit states.

Although some quantum channels proposed in some existing schemes are not suitable for CT of arbitrary $N$-qubit states from the controller's point of view, they can be used for teleporting restricted sets of states \cite{yang2,we}. For example, although we showed that the 3-qubit partially entangled MS state is ineligible for CT of arbitrary single-qubit states, that does not mean all partially entangled channels are unsuitable for CT. We can improve the control power by increasing the number of qubits Charlie has. For example, we can construct the following partially entangled channel for teleporting arbitrary single-qubit states:
\begin{eqnarray}
&&a\vert \Phi^+\rangle_{AB}\vert 00\rangle_{C}+b\vert \Phi^-\rangle_{AB}\vert 01\rangle_{C}\nonumber\\&&+c\vert \Psi^+\rangle_{AB}\vert 10\rangle_{C}+d\vert \Psi^-\rangle_{AB}\vert 11\rangle_{C}
\end{eqnarray}
where the four parameters are taken to be real and $a>b,c,d$ for simplicity. They satisfy the normalization condition $a^2+b^2+c^2+d^2=1$. After Charlie measures his two qubits in the product basis, Alice and Bob share one of four Bell states and can thus perfectly teleport a single qubit based on Charlie's measurement results.  If Charlie does not want to allow the teleportation and does not measure his qubits, then the average NCF for this channel is $a^2+1/3(b^2+c^2+d^2)$, which can made less than $2/3$ by choosing appropriate parameters for the quantum channel. We thus find that for partially entangled channels,  the control power can increase with the number of qubits held by the controller.  For teleporting $N$-qubit states, the maximum number of qubits  to be owned by the controller is  $2N$, which is equal to the information Alice has.

To summarize, we have quantitatively analyzed the controller's power in controlled teleportation schemes. We use the classical limit as a lower bound for the control power - a quantum channel is suitable for CT only if the teleportation fidelity without the permission of the controller does not exceed the classical limit. We analysed several typical CT schemes and found for teleporting arbitrary $N$-qubit states, the protocols in which each controller only deals with one qubit cannot meet the required minimum control power, while the schemes in which each controller possesses $N$ qubits can. We showed that for teleporting arbitrary $N$-qubit states, there is a trade off between the amount of quantum resources and the control power. Our criterion is simple, practical  and applicable to evaluate all CT schemes for teleporting pure states. In future work we plan to generalize our work to the case of mixed states.

\section*{Acknowledgement}

XL is supported by the National Natural Science
Foundation of China under Grant No. 11004258 and the Fundamental Research Funds for the Central Universities under
Grant No.CQDXWL-2012-014. SG acknowledges support from the Ontario Ministry of Research and Innovation and the Natural Sciences and Engineering Research Council of Canada.

\end{document}